# Observation of dressed intra-cavity dark states

Yanhua Wang[1-2], Jiepeng Zhang[3-4], and Yifu Zhu[1]

[1]Department of Physics, Florida International University, Miami, Florida 33199
[2] College of Physics and Electronics, Shanxi University, Taiyuan, 030006, China
[3]Wuhan Institute of Physics and Mathematics, Chinese Academy of Sciences, Wuhan, China
[4]Physics Division P-23, Los Alamos National Laboratory, Los Alamos, NM 87544

**Abstract:** Cavity electromagnetically induced transparency in a coherently prepared cavity-atom system is manifested as a narrow transmission peak of a weak probe laser coupled into the cavity mode. We show that with a resonant pump laser coupling the cavity-confined four-level atoms from free space, the narrow transmission peak of the cavity EIT is split into two peaks. The two peaks represent the dressed intra-cavity dark states and have a frequency separation approximately equal to the Rabi frequency of the free-space pump laser. We observed experimentally the dressed intra-cavity dark states in cold Rb atoms confined in a cavity and the experimental results agree with theoretical calculations based on a semiclassical analysis.

## 1. Introduction

Electromagnetically induced transparency (EIT) can be created in various atomic systems via coherent interactions of radiations fields and atoms, and has been shown to be important for various applications in quantum optics and nonlinear optics [1-5]. Recent studies of EIT and related phenomena have been extended to coherent coupled atom-cavity systems [6-11]. It has been shown that in a coherently coupled cavity and multi-atom system, the interplay of the collective coupling of the atoms and the cavity mode, and the atomic coherence and interference manifested by EIT may lead to interesting linear and nonlinear optical phenomena [12-13]. Recently, all-optical switching at low light intensities has been observed in a cavity-confined four-level EIT system coherently coupled by multiple laser fields [14-15]

Here we report an experimental study of an atom-cavity system consisting of N four-level atoms confined in an optical cavity and coherently coupled from free space by two laser fields: one acts as a coupling laser and forms a $\Lambda$-type standard EIT configuration with the cavity mode; another acts as a pump (dressing) laser and forms a N-type coupled atomic system with the coupling laser and the cavity mode. Without the cavity, such coherently prepared four-level EIT system (see Fig. 1(a)) has been studied before and is shown to be useful for applications such as the EIT enhanced nonlinear absorption and cross-phase modulation at low light levels [16-24]. It has been observed that in the free-space four-level EIT system, the resonant pump laser interrupts the EIT destructive interference and induces large $3^{rd}$-order nonlinear absorption [18-20]. The spectral manifestation of the enhanced nonlinearities is the appearance of the absorption peak in the EIT transmission window of a weak probe laser (see Fig. 3(b)). Here we show that with the four-level EIT system confined in a cavity, the transmission spectrum of a weak probe laser through the cavity is qualitatively different from that of the free space: without the pump laser, we observe the cavity EIT, a narrow transmission peak at the atomic resonance; when the pump laser is present, the narrow transmission peak of the cavity EIT is split into two peaks. The two peaks represent the dressed intra-cavity dark states that are produced through the combined coherent interactions of the atoms with the coupling laser, the pump laser, and the collective coupling of the cavity mode.

## 2. Theoretical analysis

We consider a composite atom-cavity system that consists of a single mode cavity containing N identical four-level atoms driven by a coupling laser and a pump (dressing) laser from free space as shown in Fig. 1(b). The cavity mode couples the atomic transition |1>-|3>. The classical coupling laser drives the atomic transition |2>-|3> with Rabi frequency $2\Omega$, and the classical pump (dressing) laser drives the atomic transition |2>-|4> with Rabi frequency $2\Omega_d$. $\Delta = \nu - \nu_{23}$ is the coupling frequency detuning, $\Delta_d = \nu_d - \nu_{24}$ is the pump (dressing) laser detuning, and $\Delta_c = \nu_c - \nu_{13}$ is the cavity-atom detuning. We calculate the transmission intensity of a weak probe laser (not shown in Fig. 1(b)) coupled into the cavity mode as the probe frequency detuning $\Delta_p = \nu_p - \nu_{13}$ is scanned across the atomic transition frequency $\nu_{13}$. For comparison, the four-level atomic system in free space and coupled by the same coupling and pump lasers is depicted in Fig. 1(a).

The interaction Hamiltonian for the cavity-atom system is

$$H = -\hbar(\sum_{i=1}^{N} \Omega \hat{\sigma}_{32}^{(i)} + \sum_{i=1}^{N} \Omega_d \hat{\sigma}_{42}^{(i)} + \sum_{i=1}^{N} g\hat{a}\hat{\sigma}_{31}^{(i)}) + H.C. \,, \qquad (1)$$

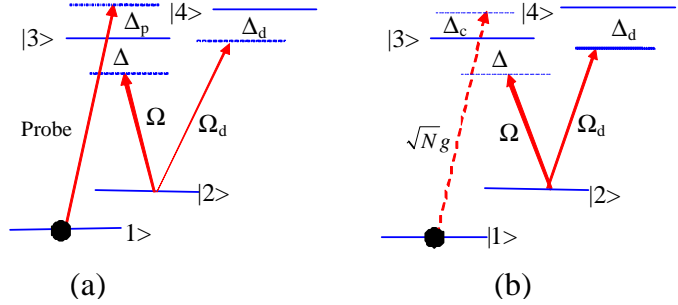

Fig. 1 (a) The energy level diagram of coherently coupled four-level atoms in free space. A coupling laser drives |2> - |3> transition with Rabi frequency $2\Omega$ and a pump laser couples |2> - |4> transition with Rabi frequency $2\Omega_d$. $\Delta$ ($\Delta_d$) is the coupling (pump) detuning. A weak probe laser couples |1> - |3> transition with a detuning $\Delta_p$. (b) The energy level diagram of coherently coupled four-level atoms in a cavity. The cavity mode is coupled to the atomic transition |1> - |3> with the collective coupling coefficient $\sqrt{N}g$ ($g = \mu\sqrt{\omega_a/2\hbar\varepsilon_0 V}$) ($\Delta_c$ is the cavity-atom detuning). The coupling laser and the pump laser are the same as in (a). Not showing is a weak probe laser coupled into the cavity mode.

where $\hat{\sigma}_{lm}^{(i)}$ (l, m=1-4) is the atomic operator for the ith atom and $\hat{a}$ is the annihilation operator of the cavity photons. The resulting operator equations of motion for the intra-cavity light field (two-sided cavity, one input) is given by [25-26]

$$\dot{\hat{a}} = -\frac{i}{\hbar}[\hat{a}, H] - \frac{\kappa_1 + \kappa_2}{2}\hat{a} + \sqrt{\kappa_1}\hat{a}_p^{in} \,, \qquad (2)$$

where $\hat{a}_p^{in}$ is the input probe field. The equation of the motion for the expectation value of the intra-cavity probe field is [14]

$$\dot{a} = -((\kappa_1 + \kappa_2)/2 - i\Delta_c)a + \sum_{i=1}^{N} ig\sigma_{31}^{(i)} + \sqrt{\kappa_1}a_p^{in} \,, \qquad (3)$$

For a symmetric cavity as in our experiment, $\kappa_1 = \kappa_2 = \kappa$. Under the EIT condition (g<<$\Omega$), the atomic population is concentrated in |1> and the steady-state solution of the intra-cavity probe field is given by

$$a = \frac{\sqrt{\kappa}a_p^{in}}{\kappa - i\Delta_c - i\chi} \,, \qquad (4)$$

where $\chi$ is the atomic susceptibity given by

$$\chi = \cfrac{ig^2 N}{\Gamma_3 - i\Delta_p + \cfrac{\Omega^2(\Gamma_4 + \gamma_{12} - i(\Delta_d + \Delta_p - \Delta))}{(\gamma_{12} - i(\Delta_p - \Delta))(\Gamma_4 + \gamma_{12} - i(\Delta_d + \Delta_p - \Delta)) + \Omega_d^2}} \quad . \tag{5}$$

The transmitted probe field is then given by $a_p^{out} = \sqrt{\kappa} a$.

Fig. 2 plots the transmitted intensity of the probe field $\dfrac{I_{out}}{I_{in}} = \dfrac{|a_p^{out}|^2}{|a_p^{in}|^2}$ versus the probe frequency detuning $\Delta p/\Gamma_3$. For simplicity, the parameters are chosen such that $\Gamma_3 = \Gamma_4 = \Gamma$, $g\sqrt{N} = 3.5\Gamma$, $\kappa = 1.5\Gamma$, $\Omega = 2\Gamma$, $\gamma_{12}=0.001\Gamma$, and $\Delta_c = \Delta = \Delta_d = 0$. Fig. 2(a) shows the probe transmission spectrum without the pump (dressing) laser ($\Omega_d$=0). The central peak at $\Delta p$=0 represents the cavity EIT, or intra-cavity dark state [8]. The two sideband peaks represent the normal modes of the coupled cavity-atom system [27-29], which are modified by the free-space coupling laser [8-9, 30]. Fig. 2(b) plots the probe spectrum with the dressing laser ($\Omega_d$=$\Gamma$), which shows that the central EIT peak is split into two peaks and the peak separation is approximately equal to $2\Omega_d$. For comparison, we plot in Fig. 3 the corresponding probe transmission spectra for the four-level system in free space (without the cavity) and coupled by the coupling laser and the pump laser with the same parameters. Without the pump laser (Fig. 3(a), the probe transmission spectrum exhibits the standard EIT spectral profile in free space (with a transparency window at the resonance $\Delta p$=0); when the pump laser is present, an absorption peak appears in the EIT window and represents the enhanced nonlinear absorption in the four-level system [17-20].

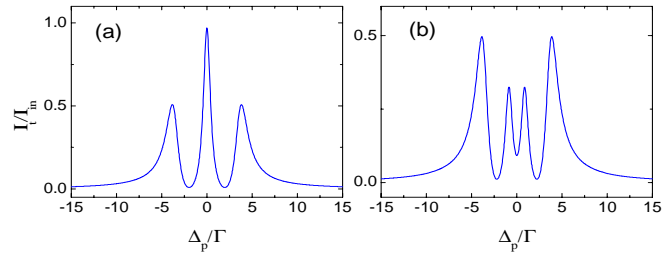

Fig. 2 The normalized transmission intensity $I_{out}/I_{in}$ of the probe laser through the cavity versus the probe detuning $\Delta_p/\Gamma$. (a) The probe transmission spectrum without the pump (dressing) laser ($\Omega_d$=0). The central peak represents the cavity EIT. (b) The probe transmission spectrum with the pump (dressing) laser ($\Omega_d$=$\Gamma$).

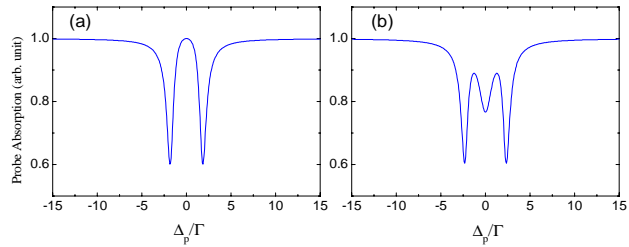

Fig. 3 The normalized transmission intensity of the probe laser through the four-level

atomic system in free space versus the probe detuning $\Delta_p/\Gamma$. (a) The probe transmission spectrum without the pump laser ($\Omega_d=0$). (b) The probe transmission spectrum with the pump laser ($\Omega_d=\Gamma$). The parameters are the same as that in Fig. 2.

## 3. Experimental results

We performed the experiment with cold $^{85}$Rb atoms confined in a near confocal cavity consisting of two mirrors of 5 cm curvature with a mirror separation ~ 5 cm. The empty cavity finesse is measured to be ~ 150. A detailed description of our experimental set up can be found in our early publications [30-31] and is briefly outlined here. Three extended-cavity diode lasers were used as the coupling laser that drives the $^{85}$Rb $D_1$ transition F=3 to F'=3, the pump (dressing) laser that couples the $^{85}$Rb $D_2$ transition F=3 to F'=4, and the probe lasers that couples the $^{85}$Rb $D_1$ transition F=2 to F'=3. The circularly-polarized coupling laser and pump laser were directed to overlap the cold atoms from the open side of the cavity and propagated in the direction nearly perpendicular to the cavity axis. The probe laser was linearly polarized parallel to the propagating direction of the coupling laser and then after sufficient attenuation, was coupled into the cavity. The transmitted probe light was collected by a photon counter (PerkinElmer SPCM-AQR-16-FC. The peak count rate of the probe light through the empty cavity is below the saturation rate of the photon counter of $10^7$ counts/s). Another part of the probe laser beam propagated nearly parallel to the coupling laser, overlapped with the cold atoms from free space, and was then collected by a photodiode, which provides the free-space absorption measurements for comparison with the cavity transmission measurements. During the measurements of the cavity transmission spectrum, the free-space part of the probe beam was blocked such that it would not interfere with the cavity transmission measurements.

The experiment was run sequentially with a repetition rate of 10 Hz. All lasers were turned on or off by acousto-optic modulators (AOM) according to the time sequence described below. For each period of 100 ms, ~98 ms was used for cooling and trapping of the $^{85}$Rb atoms, during which the trapping laser and the repump laser were turned on by two AOMs while the coupling laser, the pump laser, and the probe laser were off. The time for the data collection lasted ~ 2 ms, during which the repump laser was turned off first, and then after a delay of ~0.2 ms, the trapping laser was turned off (the current to the anti-Helmholtz coils of the MOT was always kept on), and the coupling laser, the pump laser, and the probe laser were turned on. After the coupling laser, the pump laser, and the probe laser were turned on by the AOMs for 0.2 ms, the probe laser frequency was scanned across the $^{85}$Rb $D_1$ F=2→F=3 transitions and the probe light transmitted through the cavity was then recorded versus the probe frequency detuning.

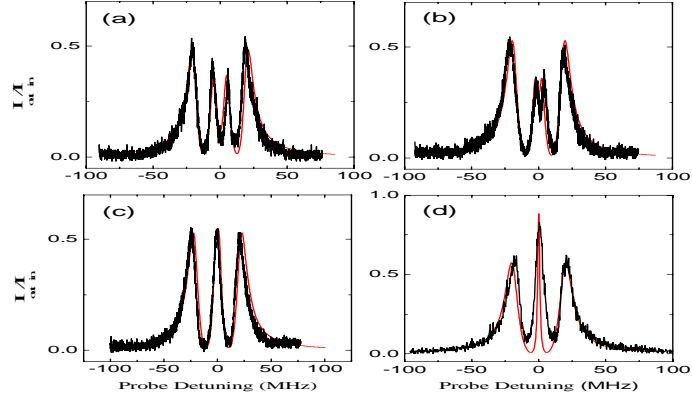

Fig. 4 The cavity transmission intensity $I_{out}/I_{in}$ versus the probe detuning $\Delta_p$. Black lines are experimental data and red lines are calculations. (a) $\Omega_d \approx 5$ MHz; (b) $\Omega_d \approx 3$ MHz; (c) $\Omega_d \approx 2$ MHz; (d) $\Omega_d = 0$.

Fig. 4 plots the measured cavity transmission intensity of the probe laser $I_{out}/I_{in}$ ($I_{in}$ is the resonant transmission of the probe light through an empty cavity) versus the probe frequency detuning $\Delta_p$. The empty cavity is tuned to the atomic transition frequency $\Delta_c = \nu_c - \nu_{13} = 0$ and both the coupling laser and the pump (dressing) laser are on resonance ($\Delta \approx 0$ and $\Delta_d \approx 0$). The decay linewidth of the Rb transitions are $\Gamma_3 = 5.7$ MHz and $\Gamma_4 = 5.9$ MHz, respectively. Other parameters are $g\sqrt{N} = 20$ MHz, $\kappa = 10$ MHz, $\Omega = 12$ MHz, $\gamma_{12} = 0.01\Gamma$, and $\Delta_c = \Delta = \Delta_d = 0$. The measured spectrum was the average of 50 scans. Fig. 4(d) shows that without the pump (dressing) laser, the three-peaked cavity EIT spectrum was observed: two sideband peaks located at $\Delta_p = \pm\sqrt{\Omega^2 + g^2 N}$ represent the normal modes of the coupled cavity-atom system, and a central peak at $\Delta_p = 0$ is manifested by EIT (the intra-cavity dark state) [8]. When the pump (dressing) laser is present, the cavity EIT peak is split into two peaks at sufficiently large $\Omega_d$ values (Fig. 4(a) and 4(b)). The splitting decreases with decreasing $\Omega_d$ values (the decreasing pump (dressing) laser intensity) and eventually disappears when $\Omega_d < 3$ MHz (Fig. 4(b) and 4(c)).

For comparison Fig. 5 plots the probe transmission spectrum through the four-level EIT system in free space. Fig. 5(b) shows the usual EIT spectrum without the pump laser: an EIT transparent window was observed at the probe resonance ($\Delta_p = 0$). With the pump laser, the transparent dip is turned into an absorption peak that represents the enhanced nonlinear absorption [16-19].

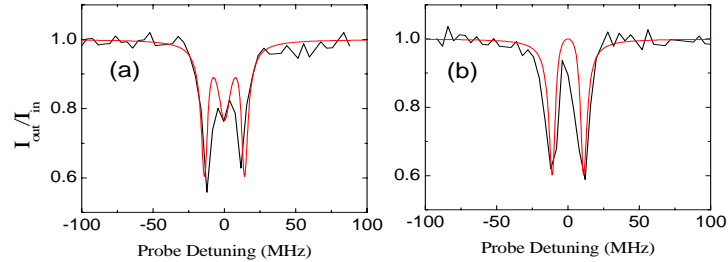

Fig. 5 The probe transmission intensity through the four-level cold Rb atoms in free space versus the probe detuning $\Delta_p$. Black lines are experimental data and red lines are calculations. (a) With the pump laser ($\Omega_d = 4.5$ MHz). (b) Without the pump laser. The other parameters are the same as that in Fig. 4(a).

In order to understand the observed spectral features in the coherent coupled four-level-atom and cavity system, we diagonalize the interaction Hamiltonian and derive the eigenvalues of the coupled cavity-atom system. Consider the resonantly coupled four-level cavity-atom system in Fig. 1(b) ($\Delta_c = \Delta = \Delta_d = 0$), The collective basis states of the atom and fields are $|1> = |1,1,..........\ .......1>|1_p>|n>|n_d>$, $|2> = \frac{1}{\sqrt{N}}\sum_{j=1}^{N}|1......2_j......1>|0_p>|n+1>|n_d>$, $|3> = \frac{1}{\sqrt{N}}\sum_{j=1}^{N}|1......3_j......1>|0_p>|n>|n_d>$, and $|4> = \frac{1}{\sqrt{N}}\sum_{j=1}^{N}|1......4_j......1>|0_p>|n+1>|n_d-1>$. Here $|1_p>(|0_p>)$ is the one (zero) photon state of the intra-cavity probe field, $|n>$ is the photon number state of the coupling field, and $|n_d>$ is the photon number state of the pump (dressing) field. We treat the coupling field and the dressing field as classical field (n>>1 and $n_d$>>1) and neglect their depletion. Then the four basis states can be rewritten as $|1> = |1,1,..........\ .......1>|1_p>$, $|2> = \frac{1}{\sqrt{N}}\sum_{j=1}^{N}|1......2_j......1>|0_p>$, $|3> = \frac{1}{\sqrt{N}}\sum_{j=1}^{N}|1......3_j......1>|0_p>$, and $|4> = \frac{1}{\sqrt{N}}\sum_{j=1}^{N}|1......4_j......1>|0_p>$. Solving the interaction Hamiltonian in the four basis states (|1> and |3> are coupled by the collective coupling coefficient $g\sqrt{N}$), we obtain the energy eigenvalues

$$\lambda = \pm\sqrt{(\Omega^2 + \Omega_d^2 + g^2N \pm \sqrt{(\Omega^2 + \Omega_d^2 + g^2N)^2 - 4g^2N\Omega_d^2})/2} , \qquad (6)$$

and the four eigen-states $\Psi_\lambda = a_\lambda|1> + b_\lambda|2> + c_\lambda|3> + d_\lambda|4>$ ($\lambda$=1-4). When the probe laser is coupled into the cavity, its transmission intensity versus the probe detuning $\Delta_p$ reveals the excitation spectrum from the ground state $|1> = |1,1,..........\ .......1>|0_p>$ to the first excited eigen-states $\Psi_\lambda$. The spectrum presents four spectral peaks corresponding to the four energy eigenvalues $\lambda$. In particular, when $\Omega_d \ll g\sqrt{N}$ (or $\Omega$), the four eigenvalues become $\lambda_{1\pm} \approx \pm\sqrt{\Omega^2 + \Omega_d^2 + g^2N} \approx \pm\sqrt{\Omega^2 + g^2N}$ with the corresponding eigenstates approximately given by $\Psi_{1\pm} \approx \frac{1}{\sqrt{2}}[|3> \mp \frac{1}{\sqrt{\Omega^2 + g^2N}}(g\sqrt{N}|1> + \Omega|2>)]$ (the modified normal modes of the coupled cavity-atom system); and $\lambda_{2\pm} \approx \pm\Omega_d\sqrt{g^2N/(\Omega^2 + \Omega_d^2 + g^2N)}$, with the corresponding eigenstates approximately given by $\Psi_{2\pm} = \frac{1}{\sqrt{2}}(|4> \pm \frac{1}{\sqrt{\Omega^2 + g^2N}}(\Omega|1> - g\sqrt{N}|2>))$, which represents the two dressed intra-cavity dark states. Therefore, when the pump (dressing) laser is not present ($\Omega_d$=0), the coupled cavity-atom system only has three first-excited engenstates: two normal modes (separated by the modified vacuum Rabi frequency $2\sqrt{\Omega^2 + g^2N}$) and an intra-cavity dark state (cavity EIT) [30]. When the pump (dressing) laser is present and couples the intra-cavity dark state to the excited state |4>, two dressed intra-cavity dark states are created with the frequency separation $2\Omega_d\sqrt{g^2N/(\Omega^2 + \Omega_d^2 + g^2N)}$.

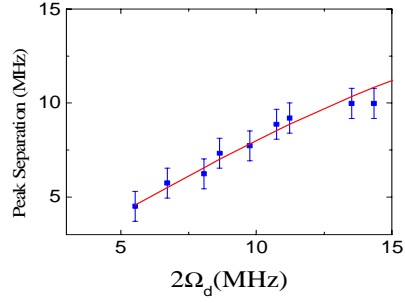

Fig. 6 The peak separation of two dressed cavity-dark states versus the Rabi frequency of the dressing laser. The parameters are $g\sqrt{N}=20$ MHz, $\kappa=10$ MHz, $\Omega=12$ MHz, and $\Delta_c=\Delta=\Delta_d=0$.

Fig. 6 plots the measured frequency splitting of the two intra-cavity dark states versus $2\Omega_d$. The solid red line is the calculated frequency separation according to Eq. (6) $\Delta'=2\lambda=2\sqrt{(\Omega^2+\Omega_d^2+g^2N-\sqrt{(\Omega^2+\Omega_d^2+g^2N)^2-4g^2N\Omega_d^2})/2}$, which agrees with the experimental measurements (filled blue squares).

## 4. Conclusion

In conclusion, we have shown that cavity EIT can be manipulated with a free-space pump laser that splits the intra-cavity dark state and creates the dressed doublet of the intra-cavity dark states. The dressed intra-cavity dark states consist of coherent superposition of the intra-cavity dark state and the excited atomic state |4>, and the frequency separation of the dressed intra-cavity dark states can be controlled by the intensity of the dressing laser. We observed the dressed intra-cavity dark states in an experiment with cold atoms confined in a cavity and coherently prepared by free-space laser fields. The experimental results agree with the theoretical calculations.

## Acknowledgement

This work is supported by the National Science Foundation under Grant No. 0757984.

References

1. S. E. Harris, Phys. Today 50, 36 (1997).

2. E. Arimondo, in *Progress in Optics*, E. Wolf ed., (Elsevier, Amsterdam, 1996) Vol. 31, p.257

3. M Fleischhauer, A. Imamoglu, & J. P. Marangos, Rev. Mod. Phys. **77**, 633-673 (2005).

4. M. D. Lukin, Rev. Mod. Phys. 75, 457(2003)

5. A. I. Lvovsky, B. C. Sanders, W. Tittel, Nature Photonics **3**, 706-714 (2009).

6. M. D. Lukin, M. Fleischhauer, M. O. Scully, and V. L. Velichansky, Opt. Lett. **23**, 295-297 (1998).

7. H. Wang, D. J. Goorskey, W. H. Burkett, and M. Xiao, Opt. Lett. **25**, 1732-1734 (2000).

8. G. Hernandez, J. Zhang, and Y. Zhu, Phys. Rev. A **76**, 053814 (2007).

9. H. Wu, J. Gea-Banacloche, and M. Xiao, Phys. Rev. Lett. **100**, 173602 (2008).

10. M. Mücke, E. Figueroa, J. Bochmann, C. Hahn, K. Murr, S. Ritter, C. J. Villas-Boas, and G. Rempe, Nature **465**, 755-758 (2010).

11. L. Slodicka, G. Hétet, S. Gerber, M. Hennrich, and R. Blatt, Phys. Rev. Lett. **105**,153604 (2010).
12. Y. Zhu, Opt. Lett. **35**, 303(2010).
13. J. Sheng, H. Wu, M. Mumba, J. Gea-Banacloche, and M. Xiao, Phys. Rev. A 83, 023829(2011).
14. M. Albert, A Dantan, and M. Drewsen, Nature Photonics 5, 633 (2011)

15. J. T. Sheng, X. H. Yang, U. Khadka, and M Xiao, Optics Express 19, 17059 (2011).

16. H. Schmidt and A. Imamoglu, Opt. Lett. **21**, 1936 (1996).

17. S. E. Harris and Y. Yamamoto, Phys. Rev. Lett. **81**, 3611 (1998).

18. M. Yan, E. Rickey, and Y. Zhu, Opt. Lett. **26**, 548-550 (2001).

19. M. Yan, E. Rickey, and Y. Zhu, Phys. Rev. A **64**, 041801 (2001).

20. D. A. Braje, V. Balic, G. Y. Yin, and S. E. Harris, Phy. Rev. A 68, 041801(R) (2003).

21. H. Kang and Y. Zhu, Phys. Rev. Lett. **91**, 93601 (2003).
22. A. B. Matsko, I. Novikova, G. R. Welch, and M. S. Zubairy, Opt. Lett. **28**, 96 (2003).
23. D. Petrosyan and G. Kurizki, Phys. Rev. A **65**, 033833(2002).
24. Z. B. Wang, K. P. Marzlin, B. C. Sanders, Phys. Rev. Lett. **97**, 063901(2006).
25. "*Quantum Noise*", C. W. Gardiner (Springer, Berlin, Heidelberg, 1991).
26. "*Quantum Optics*", D. F. Walls and G. J. Milburn (Springer-Verlag, Berlin, Heidelberg,

1994).
27. G. S. Agarwal, Phys. Rev. Lett. **53**, 1732-1735(1984).
28. M. G. Raizen, R. J. Thompson, R. J. Brecha, H. J. Kimble, and H. J. Carmichael, Phys. Rev. Lett. **63**, 240 - 243 (1989).
29. Y. Zhu, D. J. Gauthier, S. E. Morin, Q. Wu, H. J. Carmichael, and T. W. Mossberg, Phys. Rev. Lett. **64**, 2499-2503 (1990).
30. G. Hernandez, J. Zhang, and Y. Zhu, Optics Express **17**, 4798(2009).
31. J. Zhang, G. Hernandez, and Y. Zhu, Optics Express **16**, 7860(2008).